\documentclass[a4paper,11pt]{article}
\usepackage{jinstpub} 
\usepackage{lineno}
\usepackage{caption}
\usepackage{multirow}
\usepackage{booktabs}
\usepackage{subfigure}

\title{\boldmath Muon beamtest results of high-density glass scintillator tiles}

\author[a,b]{Dejing Du}
\author[a,b,1]{, Yong Liu\note{Corresponding author.}}
\author[c]{, Hua Cai}
\author[d]{, Danping Chen}
\author[a]{, Zhehao Hua}
\author[e]{, Jifeng Han}
\author[f]{, Shan Liu}
\author[a,b]{, Baohua Qi}
\author[a,b]{, Sen Qian}
\author[h]{, Jing Ren}
\author[i]{, Xinyuan Sun}
\author[g]{, Gao Tang}
\author[f]{, Dong Yang}
\author[j]{, Shenghua Yin}
\author[k]{ and Minghui Zhang}

\affiliation[a]{Institute of High Energy Physics, Chinese Academy of Sciences, \\Yuquan Road 19B, 100049 Beijing, China}
\affiliation[b]{University of Chinese Academy of Sciences, \\Yuquan Road 19A, 100049 Beijing, China}
\affiliation[c]{China Building Materials Academy, \\Guan Zhuang Dong Li 1, 100024, Beijing, China}
\affiliation[d]{Shanghai Institute of Optics and Fine Mechanics,  Chinese Academy of Sciences, \\Qinghe Road 390, 201800, Shanghai, China}
\affiliation[e]{Institute of Nuclear Science \& Technology, Sichuan University, \\Wangjiang Road 29, 610064, Chengdu, China}
\affiliation[f]{Beijing Glass Research Institute, \\Xingguang 4th Street 5, 101111, Beijing, China}
\affiliation[g]{College of Materials and Chemistry, China Jiliang University, \\Xueyuan Street 258, 310018, Hangzhou, China}
\affiliation[h]{College of Physics and Optoelectronic Engineering, Harbin Engineering University, \\Nantong Street 145, 150001, Harbin, China}
\affiliation[i]{Department of Physics, Jinggangshan University, \\Xueyuan Road 28, 343009, Ji’an, China}
\affiliation[j]{China Nuclear Nuclear Instrument CO.,LTD., \\Hongda South Road 3, 100176, Beĳing, China}
\affiliation[k]{Shanghai Institute of Ceramics, Chinese Academy of Sciences, \\Heshuo Road 585, 201899, Shanghai, China}

\emailAdd{liuyong@ihep.ac.cn}

\abstract{To achieve the physics goal of precisely measure the Higgs, Z, W bosons and the top quark, future electron-positron colliders require that their detector system has excellent jet energy resolution. One feasible technical option is the high granular calorimetery based on the particle flow algorithm (PFA). A new high-granularity hadronic calorimeter with glass scintillator tiles (GSHCAL) has been proposed, which focus on the significant improvement of hadronic energy resolution with a notable increase of the energy sampling fraction by using high-density glass scintillator tiles. The minimum ionizing particle (MIP) response of a glass scintillator tile is crucial to the hadronic calorimeter, so a dedicated beamtest setup was developed for testing the first batch of large-size glass scintillators. The maximum MIP response of the first batch of glass scintillator tiles can reach up to 107 p.e./MIP, which essentially meets the design requirements of the CEPC GSHCAL. An optical simulation model of a single glass scintillator tile has been established, and the simulation results are consistent with the beamtest results.}


\begin{document}
\maketitle
\flushbottom

\section{Introduction}
Next-generation high energy electron-positron collider experiments have been proposed including the CEPC~\cite{CEPCCDR2}, FCC-ee~\cite{FCC}, ILC~\cite{ILC} and CLIC~\cite{CLIC} for precision measurements of the Higgs, Z/W bosons as well as searches for new physics beyond the Standard Model. To fully exploit the physics potentials, the CEPC requires accurate particle identification and reconstruction of all final states from Higgs, W and Z bosons, which requires the jet energy resolution of the CEPC detector to achieve $\rm\sim30\%/\sqrt{E_{jet}(GeV)}$~\cite{CEPCCDR2}. To address this challenge, a feasible paradigm is high granular calorimetry based on the particle flow algorithm (PFA)~\cite{PFA}, which aims to determine the energy-momentum of each particle within a jet using the optimal sub-detector. PFA-oriented calorimeters with various technical options, including digital and analog readout, have been proposed and extensively studied within the CALICE collaboration~\cite{CALICE_website} over the past two decades. The analog hadronic calorimeter (AHCAL)~\cite{AHCAL_prototype_1, AHCAL_prototype_2} considered is sampling calorimeter with steel as the absorber and scintillator tiles as the sensitive detector unit. 

Based on the PFA, the CEPC calorimeter team has proposed a novel hadronic calorimeter with glass scintillator tiles (GSHCAL)~\cite{HU2024168944} to precision measurements of jets at future Higgs factories. The GSHCAL utilizes high-density glass scintillators with a high energy sampling fraction, which can significantly improve the hadronic energy resolution in the low energy region (typically below 10 GeV for major jet components at Higgs factories)~\cite{instruments6030032}. Its detector design is based on the AHCAL technique proposed in the CEPC Conceputal Design Report~\cite{CEPCCDR2}, but uses the glass scintillators instead of the plastic scintillators.

The performance of a basic detector unit, consisting of a glass scintillator tile and a silicon photomultiplier (SiPM), is crucial to the energy resolution of the GSHCAL, especially the minimum ionizing particle (MIP) response. The MIP response provides the energy scale for the energy reconstruction of the highly granular hadronic calorimeter (HCAL). This article introduces the experimental setup and MIP response results for the glass scintillator tiles in Section~\ref{sec:setup} and Section~\ref{sec:results}. Additionally, the simulation studies will be presented in Section~\ref{sec:simulation}, followed by the discussions on the test results in Section~\ref{sec:discussion} and a summary in Section ~\ref{sec:summary}.

\section{Experimental setup}
\label{sec:setup}
We developed a dedicated test system and conducted a beamtest at the European Organization for Nuclear Research (CERN). The beamtest was carried out at CERN Proton Synchrotron (PS) T9 beamline in May 2023, where the facility can provide muons, electrons and charged hadrons (up to 15 GeV/c) ~\cite{PS}. A total of 12 scintillator tiles were measured at CERN. 

\subsection{Test setup}
\begin{figure}[htpb]
    \centering 
        \subfigure[]{\includegraphics[width=5cm]{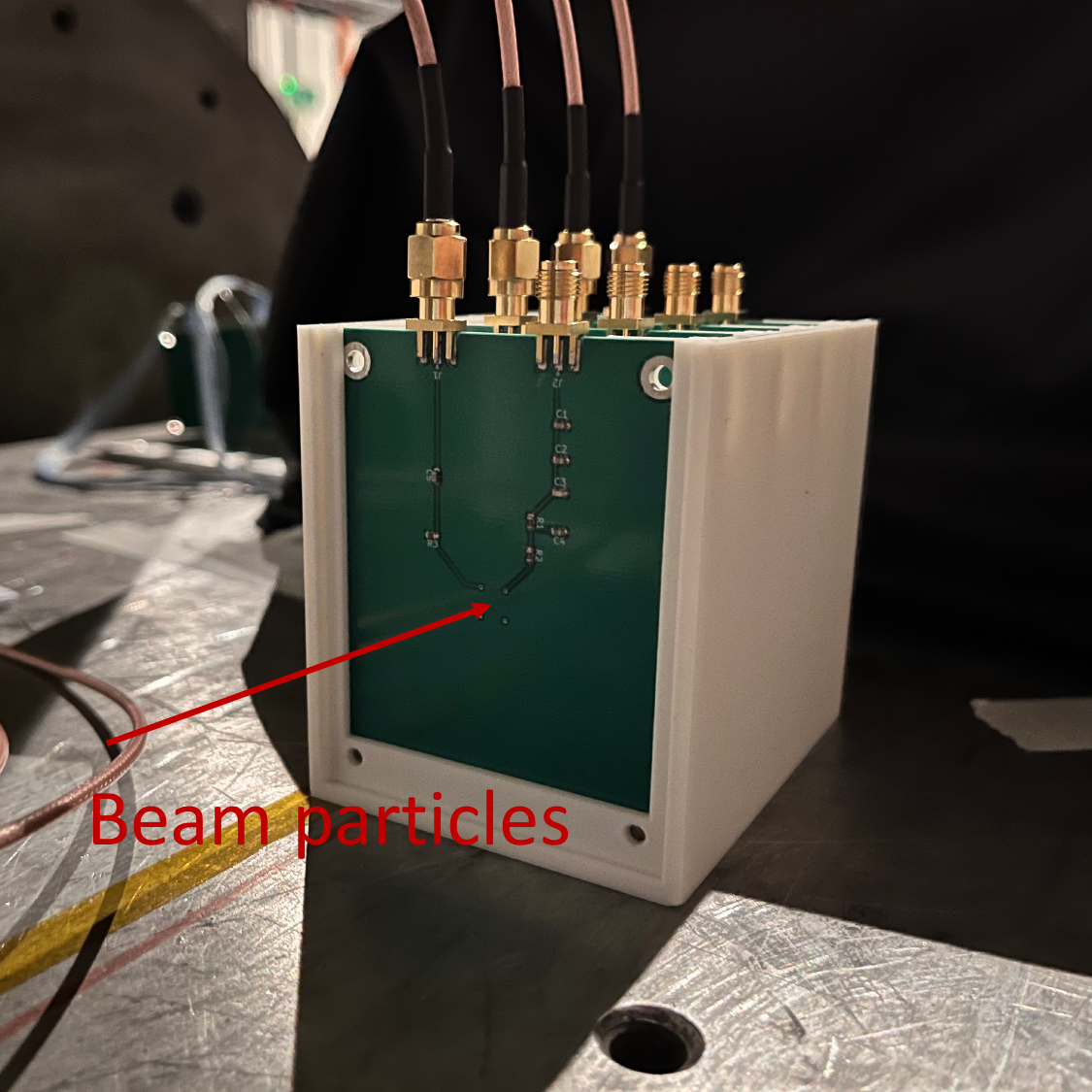}\label{fig:Setup_Photo}}
        \subfigure[]{\includegraphics[width=8cm]{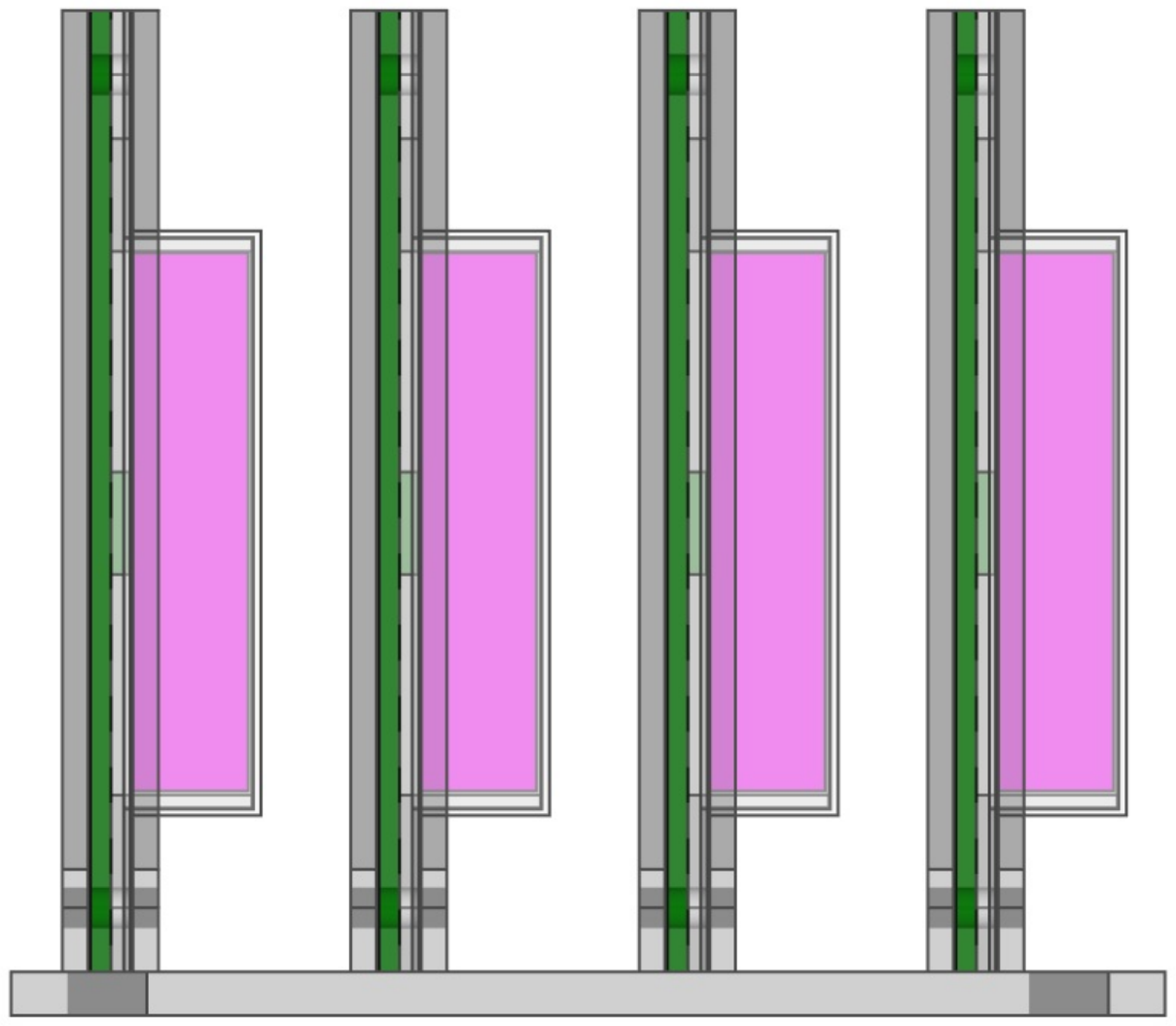}\label{fig:Setup_CAD}}
    \caption{The beamtest setup (\textbf{a}) photo and (\textbf{b}) schematic, where pink represents scintillator samples, green represents SiPM and its PCB, and gray represents 3D-printed supports.}
    \label{fig:setup}
\end{figure}

As shown in Figure~\ref{fig:setup}, 4 tiles can be simultaneously measured, comprising one plastic scintillator tile placed upstream as a reference and 3 glass scintillator tiles placed downstream. Each test sample was wrapped with reflector films and directly air-coupled with the SiPM, which is Hamamatsu's S13360-6025PE (6 $\times$ 6 mm$^{2}$ sensitive area) ~\cite{S13360-6025PE}. The signal detected by the SiPM is amplified and then acquired by a 4-channel fast oscilloscope (PicoScope 6426E). The tests use a 10 GeV muon beam with a beam spot size of $\sim$10 cm, effectively covering the entire scintillator tile.

\subsection{Scintillator glass sample}
All glass scintillator samples tested in this beamtest were developed by the Large Area Glass Scintillator Collaboration group, established in China in 2021 with a dedicated focus on key technology research and development of high-performance glass scintillators to meet the requirements of high-energy physics experiments. Based on the previous studies ~\cite{Hua_2023} for small glass samples, a batch of large-size glass scintillator samples has been developed. These glass samples includes three glass systems, namely Gd-Al-B-Si-Ce, Gd-Ba-B-Si-Ce and Gd-Ba-Al-B-Si-Ce.  Due to being the first batch of developed samples, the dimensions of these samples are inconsistent, as indicated in Table~\ref{tab1}, with widths ranging from 25.6 to 40.3 mm, and thickness within the range from 5.0 to 10.2 mm. The transmission spectra of the glass samples were measured using an ultra-violet-visible spectrophotometer (Lambda 650, PerkinElmer), and the X-ray excited luminescence (XEL) spectrum of the glass samples were tested by the spectrograph (Omni-$\lambda$ 300i, Zolix) and X-ray sources~\cite{Hua}. The transmittance at a wavelength of 400 nm are shown in Table~\ref{tab1}. All glass scintillators exhibit broadband emissions in the range of 300 to 600 nm and the peak of XEL spectra is around 390 nm, which matches the photon detection efficiency (PDE) spectrum of most common SiPMs. These samples have a density of $\sim$ 5.1 g/cm$^{3}$ and a refractive index of around 1.73.

\begin{table}[htpb]
\caption{Parameters of glass scintillators}
\centering
\label{tab1}
 	\begin{tabular}{cccccc}
    \hline
    \multirow{2}{*}{\textbf{Glass Index}} & \textbf{Transverse Size} & \textbf{Thickness} & \textbf{Transmittance} \\
     & \textbf{mm $\times$ mm} & \textbf{mm} & \textbf{\%} \\
    \hline
    \#1  & 33.5 $\times$ 27.6 & 5.1  & 69 \\
    \#2  & 30.2 $\times$ 29.5 & 6.6  & 61 \\
    \#3  & 29.9 $\times$ 28.1 & 10.2 & 70 \\
    \#4  & 37.2 $\times$ 35.1 & 5.3  & 80 \\
    \#5  & 40.0 $\times$ 35.1 & 4.2  & 78 \\
    \#6  & 40.3 $\times$ 29.8 & 9.4  & 55 \\
    \#7  & 34.8 $\times$ 34.8 & 7.5  & 65 \\
    \#8  & 27.8 $\times$ 25.6 & 5.0  & 81 \\
    \#9  & 34.7 $\times$ 34.6 & 7.5  & 49 \\
    \#10 & 35.2 $\times$ 34.7 & 7.4  & 64 \\
    \#11 & 30.5 $\times$ 30.0 & 8.7  & 81 \\
    \hline
 	\end{tabular}
\end{table}

\section{Beamtest data analysis and results}
\label{sec:results}
\begin{figure}[htbp]
    \centering 
    \includegraphics[width=10cm]{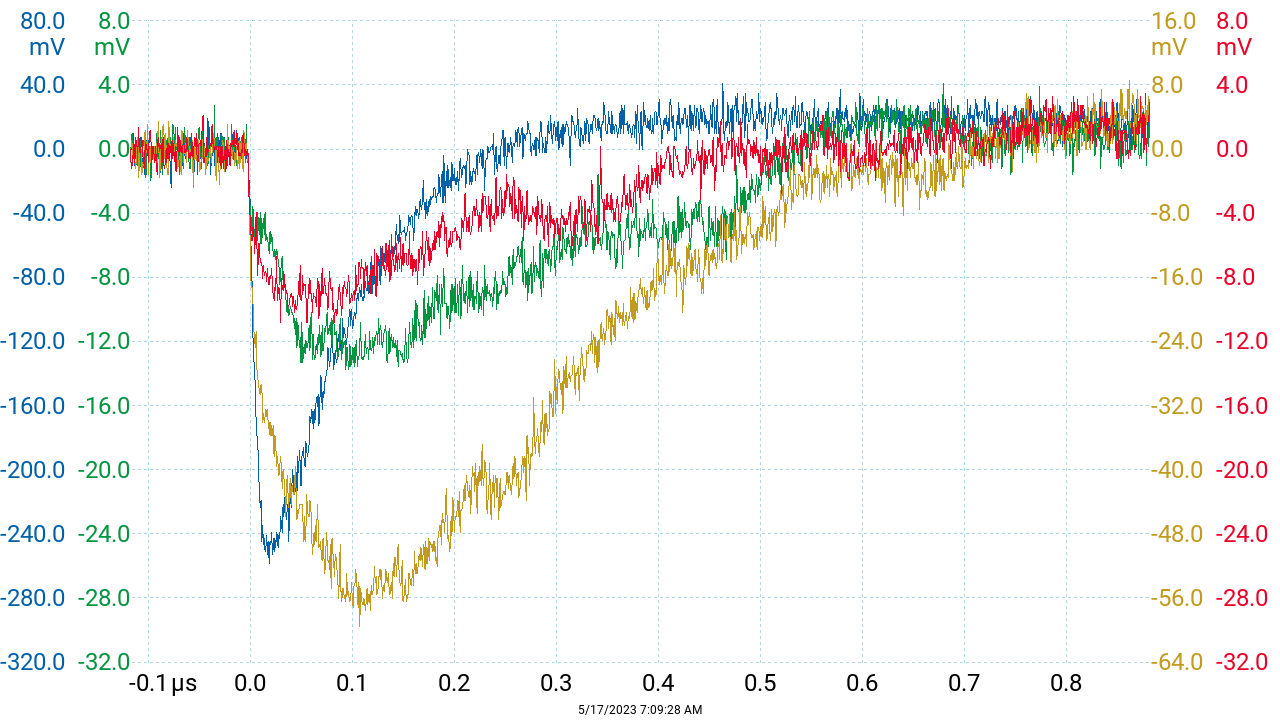}
    \caption{The typical waveforms: the blue curve corresponds to the plastic scintillator tile, while the green, yellow, and red curves correspond to the glass scintillator tiles.}
    \label{fig:waveform}
\end{figure}

\begin{figure}[htbp]
    \centering 
        \subfigure[]{\includegraphics[width=7cm]{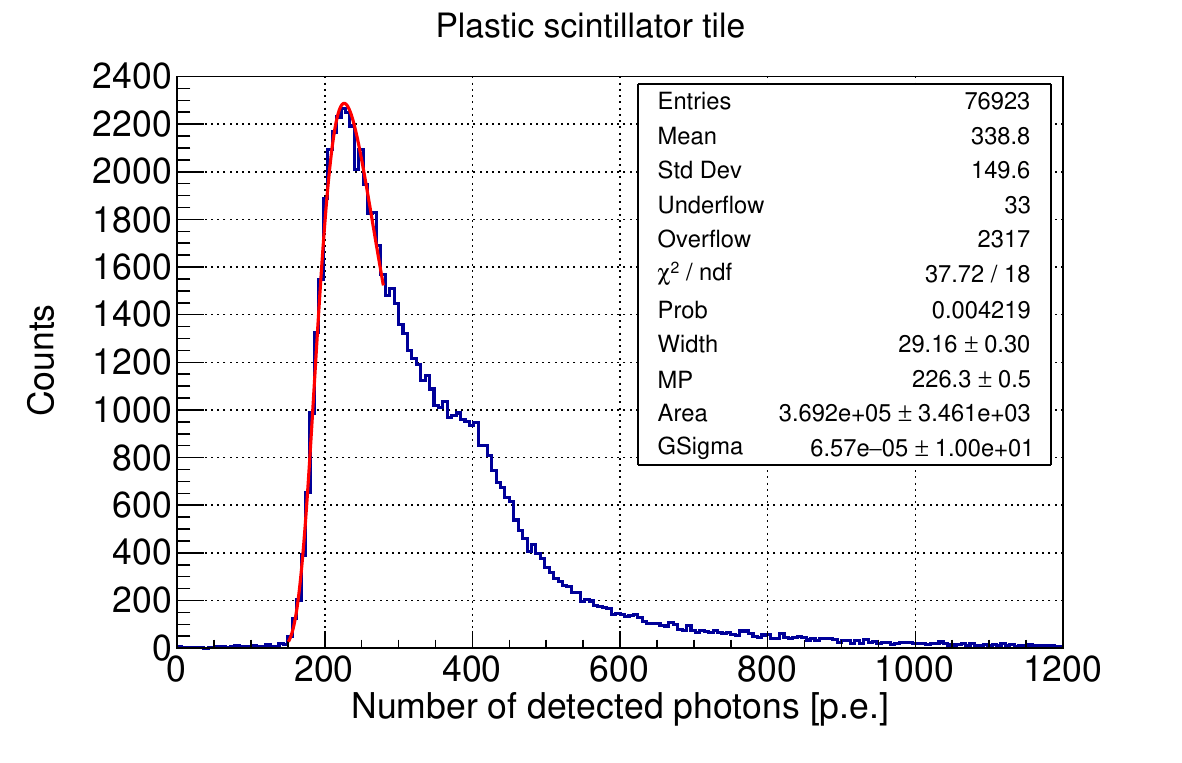}\label{fig:MIP_plastic}}
        \subfigure[]{\includegraphics[width=7cm]{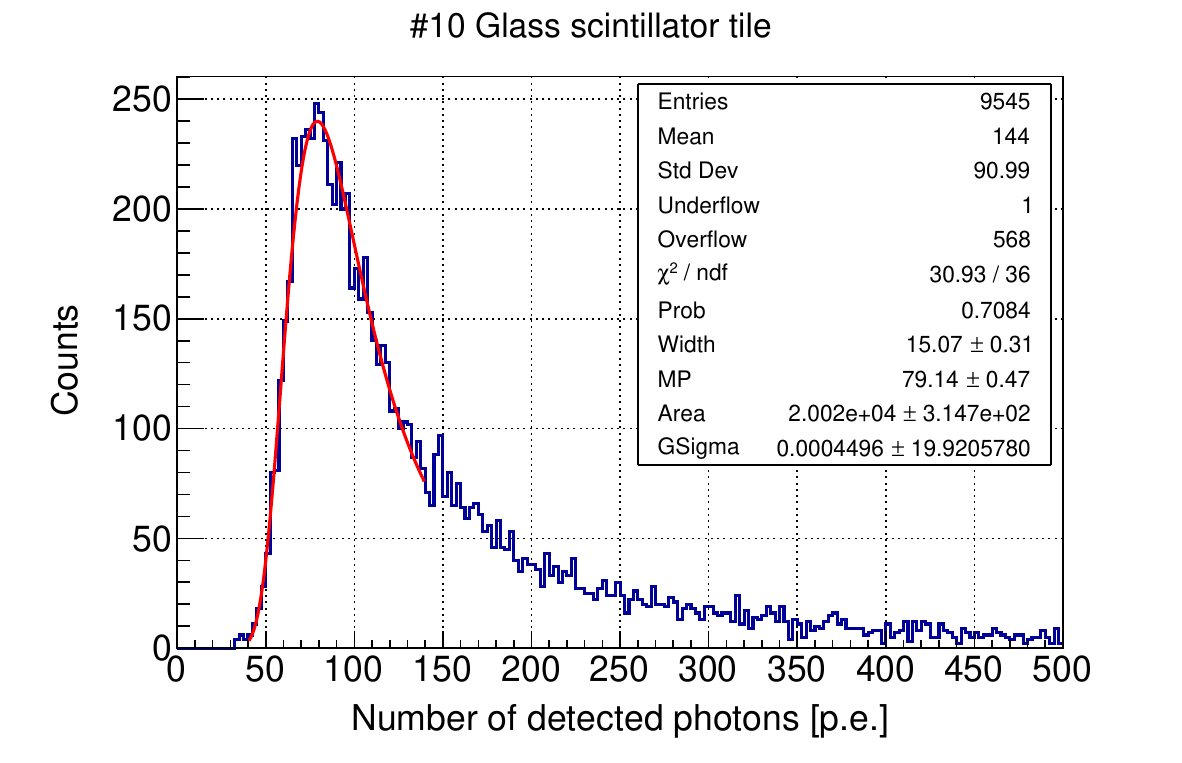}\label{fig:MIP_CERN10}}
        \subfigure[]{\includegraphics[width=7cm]{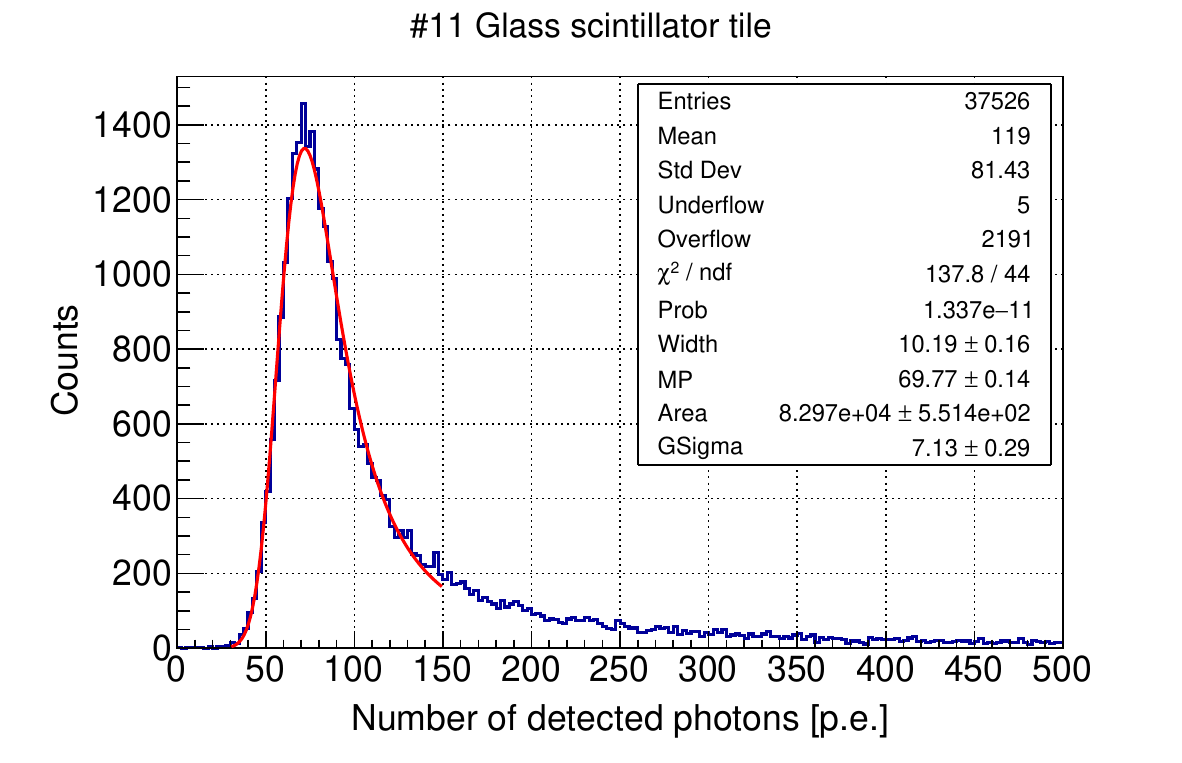}\label{fig:MIP_CERN11}}
        \subfigure[]{\includegraphics[width=7cm]{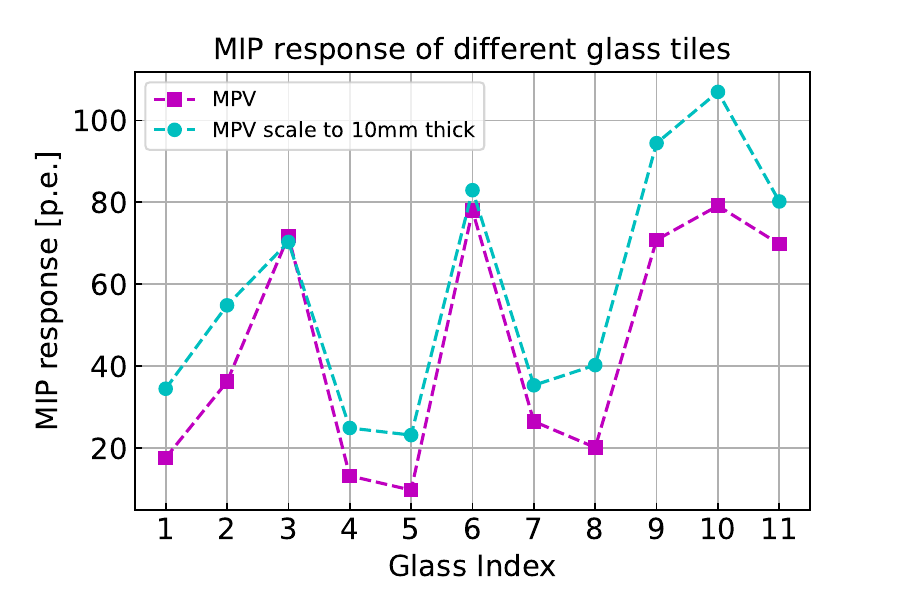}\label{fig:MIP}}
    \caption{MIP response spectra for (\textbf{a}) plastic scintillator tile, (\textbf{b}) \#10 glass scintillator tile and (\textbf{c}) \#11 glass scintillator tile. (\textbf{d}) MIP responses for all glass scintillator tiles, with magenta points representing measured values and blue points representing values scaled by 10 mm thickness.}
    \label{fig:MIPresponse}
\end{figure}

 All 11 glass scintillator tiles were successfully tested with the 10 GeV muon beam. Figure~\ref{fig:waveform} shows the typical waveforms of scintillator tiles captured by the oscilloscope. After performing waveform integration and single photoelectron calibration, the spectrum of the number of photoelectrons (p.e.) detected by each tile can be obtained, as shown in Figure~\ref{fig:MIPresponse}. The MIP response is defined as the most probable value (MPV) obtained from fitting this spectrum with a Landau convoluted Gaussian function.

Clear MIP signals were observed in all 11 glass scintillator tiles. As shown in Figure~\ref{fig:MIPresponse}, the MIP response of the plastic scintillator tile, the \#10 glass scintillator tile and the \#11 glass scintillator tile are 226, 79 and 70 p.e./MIP, respectively. Figure~\ref{fig:MIP} shows the MIP response of different glass scintillator tiles, and the magenta points represent the measured values for each glass tiles. To compare samples of different thicknesses, assuming that the MIP response is proportional to the thickness in the ideal case, we scaled the MIP response linearly to a 10 mm thickness, which is the baseline design of CEPC GSHCAL tiles. As shown by the blue points in Figure \ref{fig:MIP}, the scaled MIP responses of all glass samples range from 23 to 107 p.e./MIP, which is close to the target value for the CEPC GSHCAL. In the current research status, the GSHCAL requires the MIP response of the basic detector unit to achieve 100-150 p.e./MIP with a glass scintillator tile thickness of 10 mm.

\begin{figure}[htbp]
    \centering 
        \subfigure[]{\includegraphics[width=7cm]{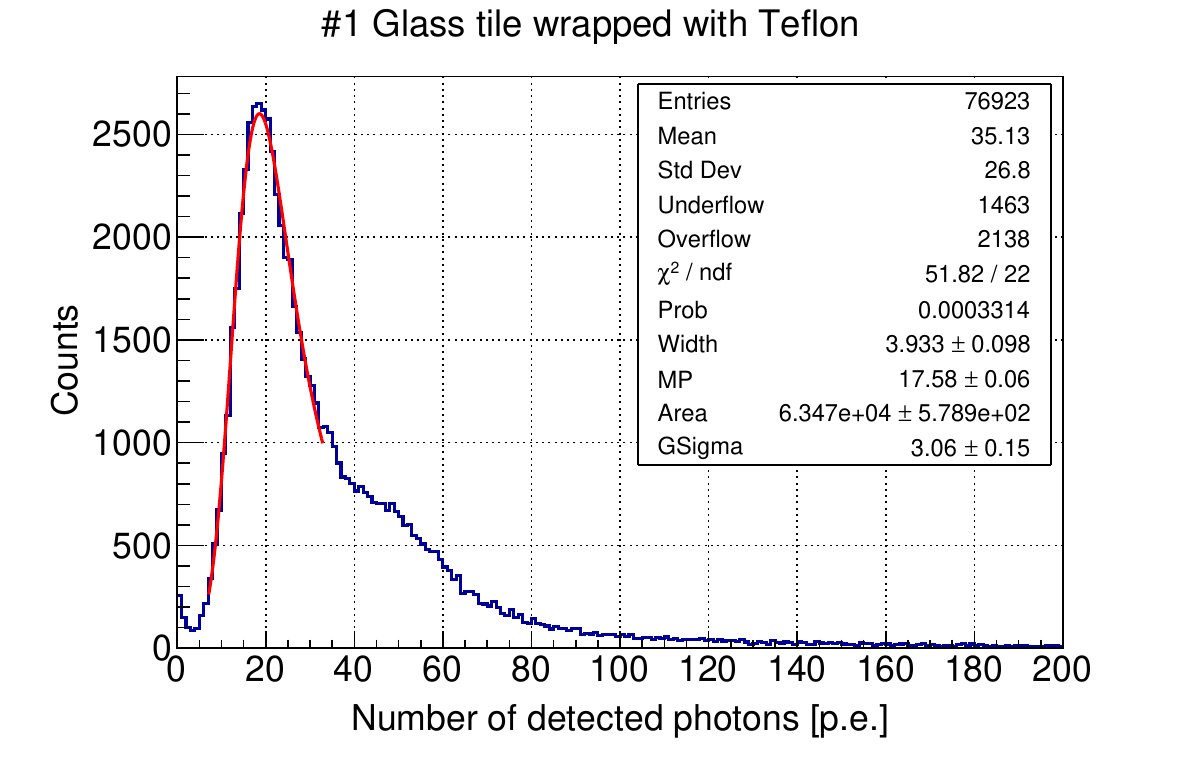}\label{fig:MIP_CERN1}}
        \subfigure[]{\includegraphics[width=7cm]{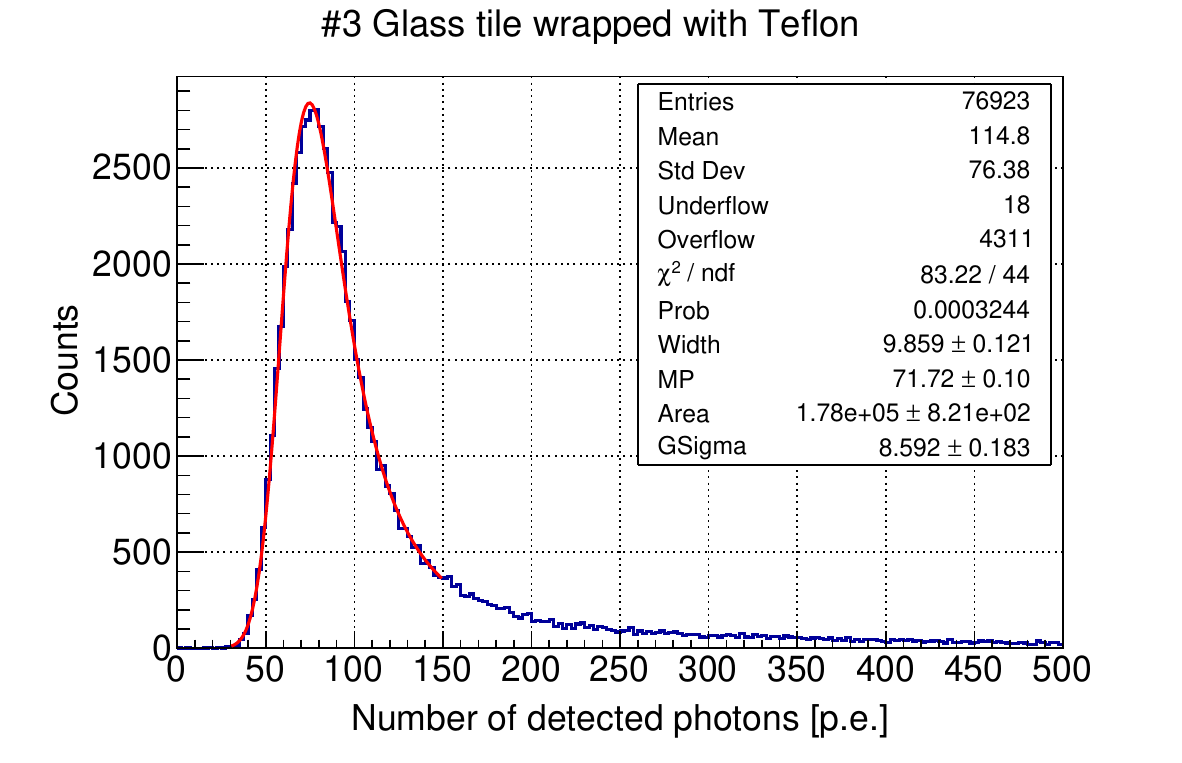}\label{fig:MIP_CERN3}}
        \subfigure[]{\includegraphics[width=7cm]{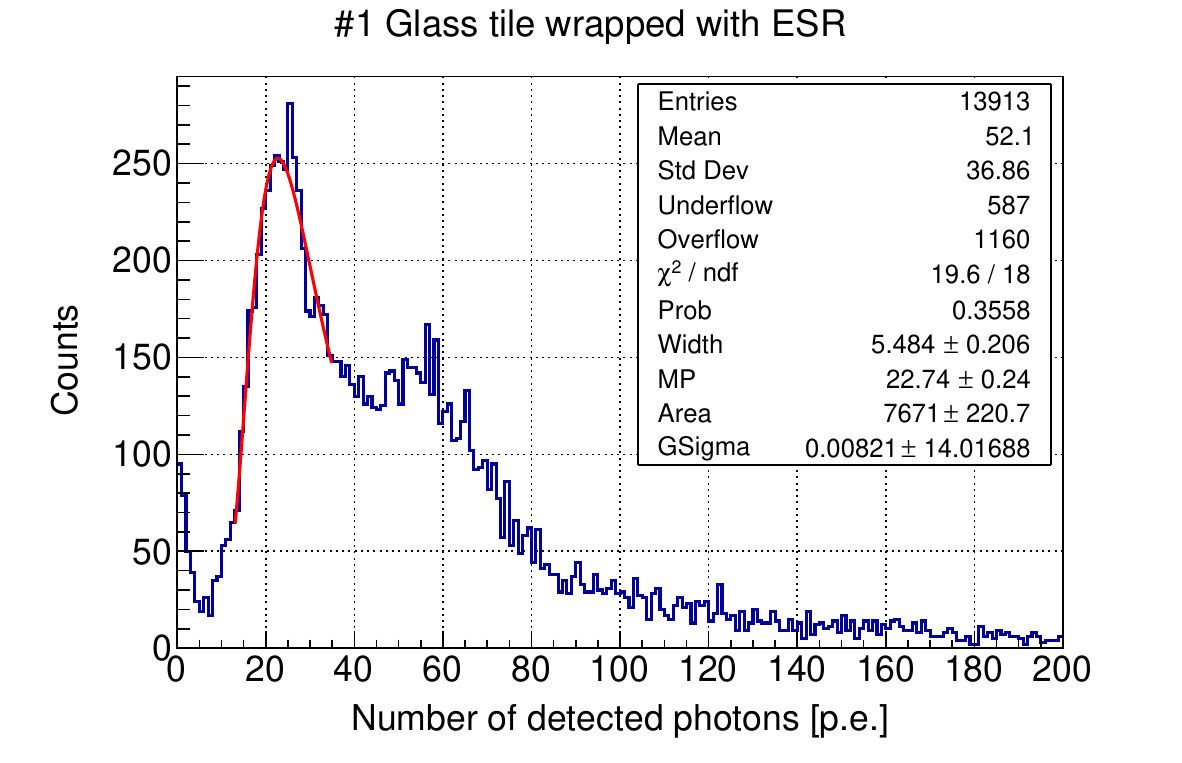}\label{fig:MIP_CERN1_ESR}}
        \subfigure[]{\includegraphics[width=7cm]{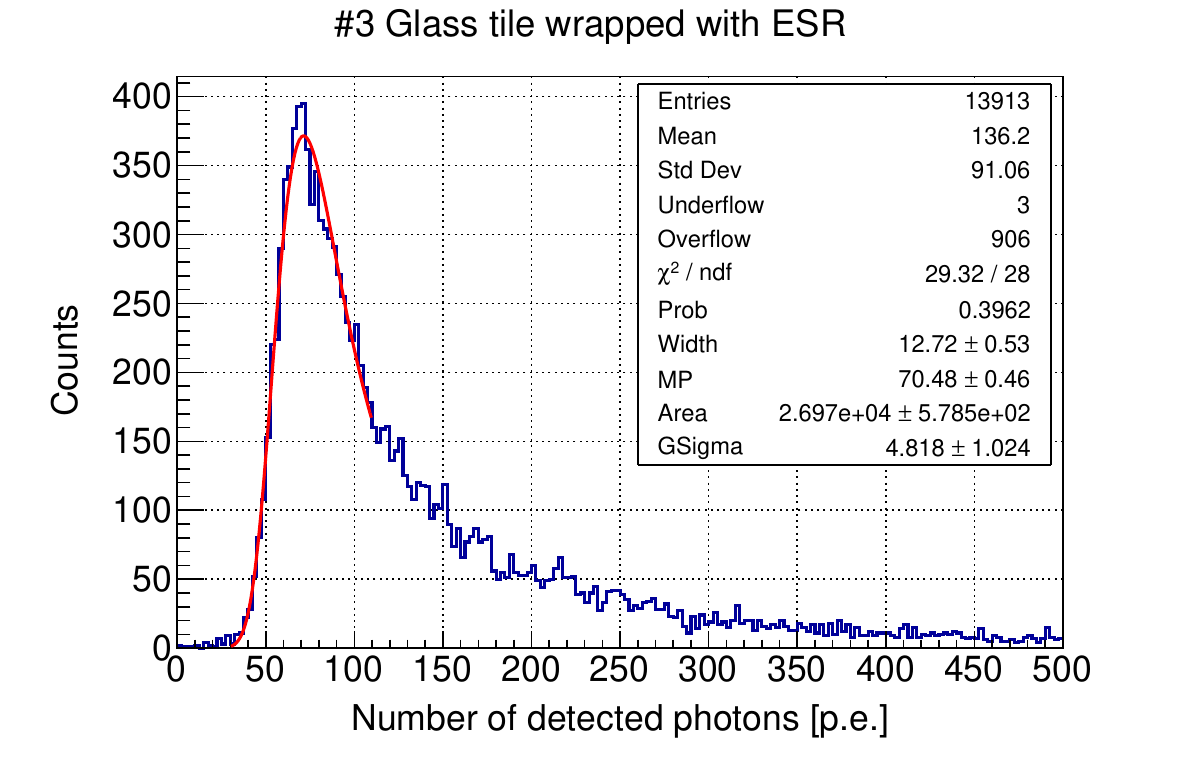}\label{fig:MIP_CERN3_ESR}}
    \caption{MIP response spectra for (\textbf{a}) \#1 glass wrapped with Teflon, (\textbf{b}) \#3 glass wrapped with Teflon, (\textbf{c}) \#1 glass wrapped with ESR, and (\textbf{d}) \#3 glass wrapped with ESR.}
    \label{fig:DifFilm}
\end{figure}

The impact of different reflector films on the MIP response was also measured. Two glass scintillator tiles (\#1 and \#3) were rewrapped with ESR film and tested, while keeping the test setup fixed. As shown in Figure~\ref{fig:DifFilm}, the MIP response of the glass scintillator tiles measured using Teflon and ESR is consistent in general.

Additionally, as shown in Figure~\ref{fig:MIP_plastic}, ~\ref{fig:MIP_CERN1} and ~\ref{fig:MIP_CERN1_ESR}, an unexpected structure was observed on the right side of the MIP peak in the MIP energy spectrum. These structures are caused by the non-uniformity of the scintillator tile itself and the distribution of the beam, which will be explained in the next section of the simulation studies.

\section{Simulation studies}
\label{sec:simulation}
To comprehend the peculiar structures observed in the MIP spectra from the beamtest, a Geant4 ~\cite{Geant4} optical simulation (with version 10.7.4 and the physics list “QGSP\_BERT”) has been established for the scintillator tiles. All setups in the optical simulation are consistent with the beamtest, taking into account light production and transport in the scintillator tiles, as well as the SiPM PDE. The primary particle beam consists of monochromatic muons with an energy of 10 GeV. The distribution of the particles in the transverse plane is a two-dimensional Gaussian distribution ($\sigma$ = 5 mm). Due to the constraints of the trigger, only particles impinging in the central region of $4\times4~cm^{2}$ are considered in the analysis. Under these settings, the simulation can reproduce the beamtest results. 

\begin{figure}[htbp]
    \centering 
        \subfigure[]{\includegraphics[width=7cm]{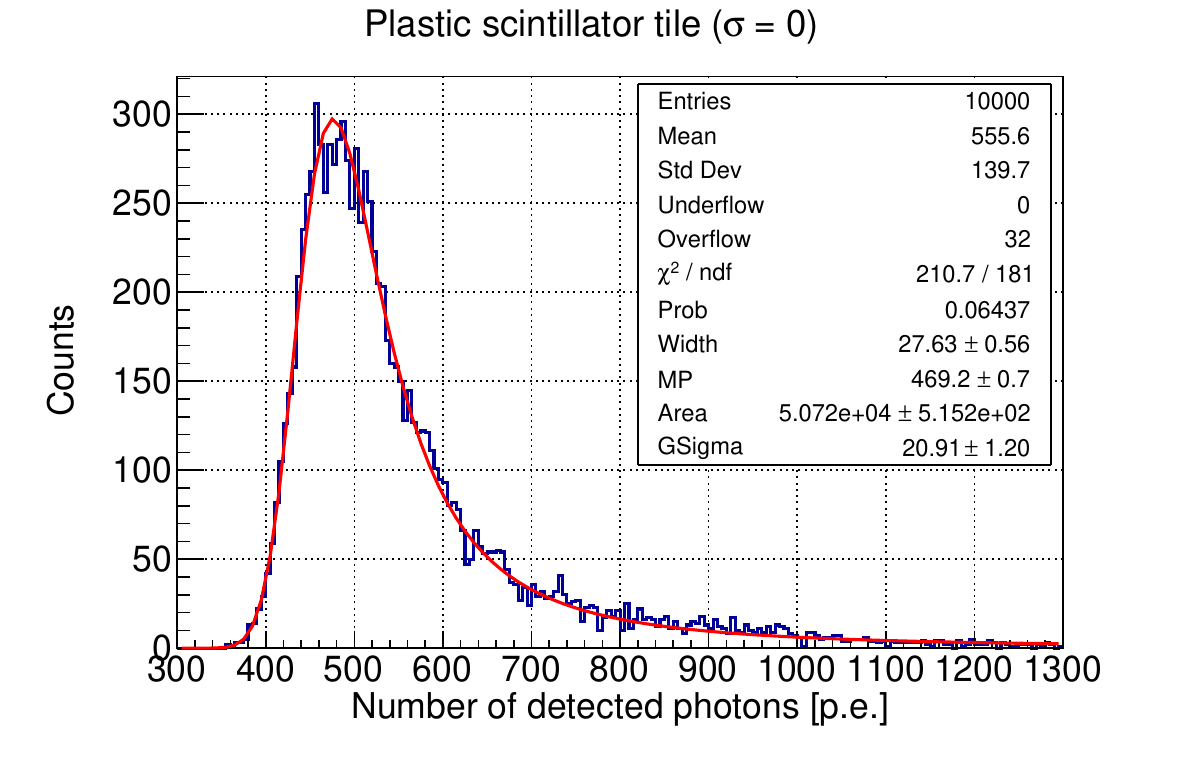}\label{fig:Sim_Plastic_Teflon_Gaus0mm}}
        \subfigure[]{\includegraphics[width=7cm]{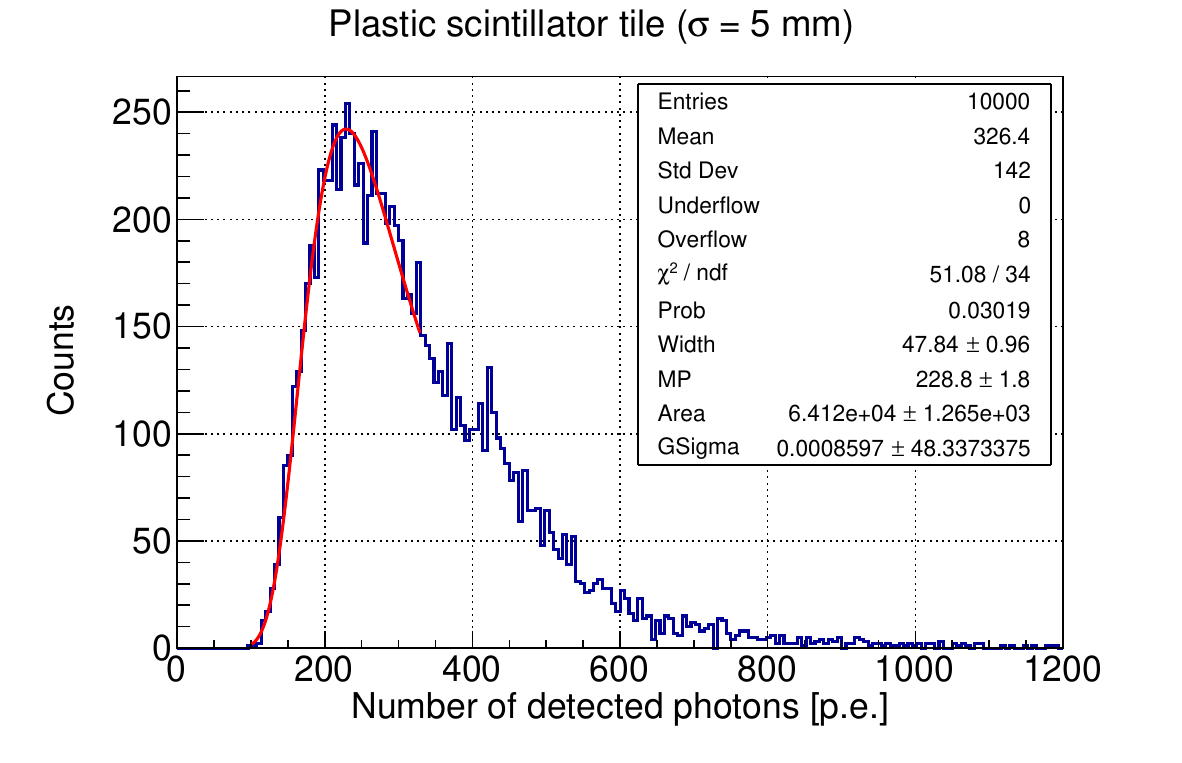}\label{fig:Sim_Plastic_Teflon_Gaus5mm}}
    \caption{Optical simulation results of MIP response for plastic scintillator tiles wrapped with Teflon film: (\textbf{a}) muons are incident perpendicularly at the center of the tile, (\textbf{b}) muon beam spot is a two dimensional Gaussian distribute ($\sigma$ = 5 mm).}
    \label{fig:Sim_Teflon}
\end{figure}

For the MIP response of the plastic scintillator tile, the value obtained from the simulation in Figure~\ref{fig:Sim_Plastic_Teflon_Gaus5mm} is 229 p.e./MIP, and the measured value in Figure~\ref{fig:MIP_plastic} is 226 p.e./MIP. The simulation and experimental results align closely, with a deviation of less than 2\%. By comparing Figure~\ref{fig:Sim_Plastic_Teflon_Gaus0mm} and Figure~\ref{fig:Sim_Plastic_Teflon_Gaus5mm}, it is evident that a Gaussian distribution of the beam spot leads to a leftward shift in the peak position of the MIP spectrum, with additional structures emerging on the right side of the peak.

\section{Discussions}
\label{sec:discussion}
Figure \ref{fig:MIP} shows MIP response of all 11 the glass samples, with the highest response reaching 107 p.e./MIP and the lowest response only at 23 p.e./MIP. The variation in responses among the 11 samples can be attributed to differences in their components, molar fractions, and preparation processes, such as melting, quenching, and deoxidization techniques. These variations could affect the scintillation light yield and other optical properties of the glass scintillator samples, thereby influencing the MIP response.

As shown in Figure~\ref{fig:Sim_Teflon}, the MPV obtained from particles incident at the center is higher than that obtained from particles incident with a Gaussian distribution. The MIP response reaches its highest when the incident position is at the geometric center of the scintillator tile, as the SiPM is coupled in the geometric center of the scintillator tile and has a much higher light collection efficiency~\cite{instruments6030032}. When the incident particle positions follow a Gaussian distribution, the majority of particles will incident positions far from the SiPM, leading to a significant part of scintillation photons being self-absorbed by the scintillator. At the same time, it is also found that when the position of incident particles is at the center of scintillator tile, the MIP response obtained using different reflective films is consistent, as most photons are directly detected without significant reflection. However, when the incident particle position follows a Gaussian distribution, the MIP response obtained using ESR film is higher than that of Teflon film. This is because the generation position of scintillation light is far from the SiPM and requires reflection and attenuation processes before being detected, in which ESR is evidently more effective for light collection than Teflon.

In fact, the impact of the particle incident position on the MIP response is the uniformity of the scintillator tile. The results indicate that the current glass scintillator samples and tile design exhibit poor uniformity and need further improvement. This can be achieved by developing glass scintillators with higher light yield and transparency, as well as optimizing the tile design through validated simulation.

\section{Summary and prospects}
\label{sec:summary}
The glass scintillator is a very promising candidate for the CEPC HCAL, so a dedicated beamtest setup was developed and used at CERN to test the first batch of large-size glass scintillator. All 11 glass scintillator samples successfully completed the beamtests and observed distinct MIP signals. The MIP responses linearly scaled to 10 mm of these samples ranged from 23 to 107 p.e./MIP, with the MIP response of the best sample essentially meeting the requirements proposed for the CEPC GSHCAL. It should be noted that the actual size of the glass tile is currently smaller than the design requirements of CEPC GSHCAL, and another key indicator, density, also falls short of the requirements. The R\&D of glass scintillator with larger size and higher density is ongoing. Furthermore, the simulation can reproduce the results of the beam test, and simulation studies shows that the additional structures in the MIP spectrum are caused by the non-uniformity of the tile. Next, we can use the validated simulation to optimize the design of the tile to improve uniformity.

\acknowledgments
The authors would like to thank the CEPC calorimeter group for the helpful discussions on the data analysis, the technical support from CERN PS-SPS beamtest facilities and the CALICE collaboration, the efforts of the Large Area Glass Scintillator Collaboration Group in glass research and development, as well as the Beijing High Energy Kedi Science and Technology Company for providing the plastic scintillator. This work receives funding support from the European Union's Horizon Europe Research and Innovation programme under Grant Agreement No 101057511 (EURO-LABS).


\bibliographystyle{JHEP}
\bibliography{main.bib}

\end{document}